
\def\doublespace{\baselineskip=20pt plus 2pt\lineskip=3pt minus
     1pt\lineskiplimit=2pt}

\def\np{\vfill\eject}
\def\singlespace{\normalbaselines}

\parindent=20pt

\def\nonarrower{\advance\leftskip by-\parindent\advance\rightskip
by-\parindent}

\def\undertext#1{$\underline{\smash{\hbox{#1}}}$}

\def\boxit#1{\vbox{\hrule\hbox{\vrule\kern3pt
	\vbox{\kern3pt#1\kern3pt}\kern3pt\vrule}\hrule}}
\def\gtorder{\mathrel{\raise.3ex\hbox{$>$}\mkern-14mu
             \lower0.6ex\hbox{$\sim$}}}
\def\ltorder{\mathrel{\raise.3ex\hbox{$<$}\mkern-14mu
             \lower0.6ex\hbox{$\sim$}}}
\def\dalemb#1#2{{\vbox{\hrule height.#2pt
	\hbox{\vrule width.#2pt height#1pt \kern#1pt
		\vrule width.#2pt}
	\hrule height.#2pt}}}

\def\hence{\hbox{{\bf .}
		\raise 7pt
	\hbox{{\bf .}
		\lower 7pt
	\hbox{{\bf .}%

	}}}}
%
\expandafter\ifx\csname phyzzx\endcsname\relax
 \message{It is better to use PHYZZX format than to
          \string\input\space PHYZZX}\else
 \wlog{PHYZZX macros are already loaded and are not
          \string\input\space again}%
 \endinput \fi
\catcode`\@=11 
\let\rel@x=\relax
\let\n@expand=\relax
\def\pr@tect{\let\n@expand=\noexpand}
\let\protect=\pr@tect
\let\gl@bal=\global
%
%
%
\newfam\cpfam
\newdimen\b@gheight             \b@gheight=12pt
\newcount\f@ntkey               \f@ntkey=0
\def\f@m{\afterassignment\samef@nt\f@ntkey=}
\def\samef@nt{\fam=\f@ntkey \the\textfont\f@ntkey\rel@x}
\def\setstr@t{\setbox\strutbox=\hbox{\vrule height 0.85\b@gheight
                                depth 0.35\b@gheight width\z@ }}
%
%
%
%
%

\font\fourteenrm  =cmr10 scaled\magstep2
\font\twelverm    =cmr12
\font\ninerm      =cmr9
\font\sixrm       =cmr6

\font\fourteenbf  =cmbx10 scaled\magstep2
\font\twelvebf    =cmbx12
\font\ninebf      =cmbx9
\font\sixbf       =cmbx6
\font\seventeeni  =cmmi10 scaled\magstep3    \skewchar\seventeeni='177
\font\fourteeni   =cmmi10 scaled\magstep2     \skewchar\fourteeni='177
\font\twelvei     =cmmi12                       \skewchar\twelvei='177
\font\ninei       =cmmi9                          \skewchar\ninei='177
\font\sixi        =cmmi6                           \skewchar\sixi='177
\font\seventeensy =cmsy10 scaled\magstep3    \skewchar\seventeensy='60
\font\fourteensy  =cmsy10 scaled\magstep2     \skewchar\fourteensy='60
\font\twelvesy    =cmsy10 scaled\magstep1       \skewchar\twelvesy='60
\font\ninesy      =cmsy9                          \skewchar\ninesy='60
\font\sixsy       =cmsy6                           \skewchar\sixsy='60

\font\fourteenex  =cmex10 scaled\magstep2
\font\twelveex    =cmex10 scaled\magstep1

\font\fourteensl  =cmsl10 scaled\magstep2
\font\twelvesl    =cmsl12
\font\ninesl      =cmsl9

\font\fourteenit  =cmti10 scaled\magstep2
\font\twelveit    =cmti12
\font\nineit      =cmti9
\font\fourteentt  =cmtt10 scaled\magstep2
\font\twelvett    =cmtt12
\font\fourteencp  =cmcsc10 scaled\magstep2
\font\twelvecp    =cmcsc10 scaled\magstep1
\font\tencp       =cmcsc10
%
%
\def\fourteenf@nts{\relax
    \textfont0=\fourteenrm          \scriptfont0=\tenrm
      \scriptscriptfont0=\sevenrm
    \textfont1=\fourteeni           \scriptfont1=\teni
      \scriptscriptfont1=\seveni
    \textfont2=\fourteensy          \scriptfont2=\tensy
      \scriptscriptfont2=\sevensy
    \textfont3=\fourteenex          \scriptfont3=\twelveex
      \scriptscriptfont3=\tenex
    \textfont\itfam=\fourteenit     \scriptfont\itfam=\tenit
    \textfont\slfam=\fourteensl     \scriptfont\slfam=\tensl
    \textfont\bffam=\fourteenbf     \scriptfont\bffam=\tenbf
      \scriptscriptfont\bffam=\sevenbf
    \textfont\ttfam=\fourteentt
    \textfont\cpfam=\fourteencp }
\def\twelvef@nts{\relax
    \textfont0=\twelverm          \scriptfont0=\ninerm
      \scriptscriptfont0=\sixrm
    \textfont1=\twelvei           \scriptfont1=\ninei
      \scriptscriptfont1=\sixi
    \textfont2=\twelvesy           \scriptfont2=\ninesy
      \scriptscriptfont2=\sixsy
    \textfont3=\twelveex          \scriptfont3=\tenex
      \scriptscriptfont3=\tenex
    \textfont\itfam=\twelveit     \scriptfont\itfam=\nineit
    \textfont\slfam=\twelvesl     \scriptfont\slfam=\ninesl
    \textfont\bffam=\twelvebf     \scriptfont\bffam=\ninebf
      \scriptscriptfont\bffam=\sixbf
    \textfont\ttfam=\twelvett
    \textfont\cpfam=\twelvecp }
\def\tenf@nts{\relax
    \textfont0=\tenrm          \scriptfont0=\sevenrm
      \scriptscriptfont0=\fiverm
    \textfont1=\teni           \scriptfont1=\seveni
      \scriptscriptfont1=\fivei
    \textfont2=\tensy          \scriptfont2=\sevensy
      \scriptscriptfont2=\fivesy
    \textfont3=\tenex          \scriptfont3=\tenex
      \scriptscriptfont3=\tenex
    \textfont\itfam=\tenit     \scriptfont\itfam=\seveni  
    \textfont\slfam=\tensl     \scriptfont\slfam=\sevenrm 
    \textfont\bffam=\tenbf     \scriptfont\bffam=\sevenbf
      \scriptscriptfont\bffam=\fivebf
    \textfont\ttfam=\tentt
    \textfont\cpfam=\tencp }
%
%

%
\def\rm{\n@expand\f@m0 }
\def\mit{\n@expand\f@m1 }         
\def\cal{\n@expand\f@m2 }
\def\it{\n@expand\f@m\itfam}
\def\sl{\n@expand\f@m\slfam}
\def\bf{\n@expand\f@m\bffam}
\def\tt{\n@expand\f@m\ttfam}
\def\caps{\n@expand\f@m\cpfam}    
\def\em@{\rel@x\ifnum\f@ntkey=0 \it \else
        \ifnum\f@ntkey=\bffam \it \else \rm \fi \fi }
\def\em{\n@expand\em@}
\def\fourteenpoint{\fourteenf@nts \samef@nt \b@gheight=14pt \setstr@t }
\def\twelvepoint{\twelvef@nts \samef@nt \b@gheight=12pt \setstr@t }
\def\tenpoint{\tenf@nts \samef@nt \b@gheight=10pt \setstr@t }
\normalbaselineskip = 19.2pt plus 0.2pt minus 0.1pt 
\normallineskip = 1.5pt plus 0.1pt minus 0.1pt
\normallineskiplimit = 1.5pt
\newskip\normaldisplayskip
\normaldisplayskip = 14.4pt plus 3.6pt minus 10.0pt 
\newskip\normaldispshortskip
\normaldispshortskip = 6pt plus 5pt
\newskip\normalparskip
\normalparskip = 6pt plus 2pt minus 1pt
\newskip\skipregister
\skipregister = 5pt plus 2pt minus 1.5pt
\newif\ifsingl@
\newif\ifdoubl@
\newif\iftwelv@  \twelv@true
\def\singlespace{\singl@true\doubl@false\spaces@t}
\def\doublespace{\singl@false\doubl@true\spaces@t}
\def\normalspace{\singl@false\doubl@false\spaces@t}
\def\Tenpoint{\tenpoint\twelv@false\spaces@t}
\def\Twelvepoint{\twelvepoint\twelv@true\spaces@t}
\def\spaces@t{\rel@x
      \iftwelv@ \ifsingl@\subspaces@t3:4;\else\subspaces@t1:1;\fi
       \else \ifsingl@\subspaces@t3:5;\else\subspaces@t4:5;\fi \fi
      \ifdoubl@ \multiply\baselineskip by 5
         \divide\baselineskip by 4 \fi }
\def\subspaces@t#1:#2;{
      \baselineskip = \normalbaselineskip
      \multiply\baselineskip by #1 \divide\baselineskip by #2
      \lineskip = \normallineskip
      \multiply\lineskip by #1 \divide\lineskip by #2
      \lineskiplimit = \normallineskiplimit
      \multiply\lineskiplimit by #1 \divide\lineskiplimit by #2
      \parskip = \normalparskip
      \multiply\parskip by #1 \divide\parskip by #2
      \abovedisplayskip = \normaldisplayskip
      \multiply\abovedisplayskip by #1 \divide\abovedisplayskip by #2
      \belowdisplayskip = \abovedisplayskip
      \abovedisplayshortskip = \normaldispshortskip
      \multiply\abovedisplayshortskip by #1
        \divide\abovedisplayshortskip by #2
      \belowdisplayshortskip = \abovedisplayshortskip
      \advance\belowdisplayshortskip by \belowdisplayskip
      \divide\belowdisplayshortskip by 2
      \smallskipamount = \skipregister
      \multiply\smallskipamount by #1 \divide\smallskipamount by #2
      \medskipamount = \smallskipamount \multiply\medskipamount by 2
      \bigskipamount = \smallskipamount \multiply\bigskipamount by 4 }
\def\normalbaselines{ \baselineskip=\normalbaselineskip
   \lineskip=\normallineskip \lineskiplimit=\normallineskip
   \iftwelv@\else \multiply\baselineskip by 4 \divide\baselineskip by 5
     \multiply\lineskiplimit by 4 \divide\lineskiplimit by 5
     \multiply\lineskip by 4 \divide\lineskip by 5 \fi }
\Twelvepoint  
\interlinepenalty=50
\interfootnotelinepenalty=5000
\predisplaypenalty=9000
\postdisplaypenalty=500
\hfuzz=1pt
\vfuzz=0.2pt
\newdimen\HOFFSET  \HOFFSET=0pt
\newdimen\VOFFSET  \VOFFSET=0pt
\newdimen\HSWING   \HSWING=0pt
\dimen\footins=8in
%
%
%
\newskip\pagebottomfiller
\pagebottomfiller=\z@ plus \z@ minus \z@
\def\pagecontents{
   \ifvoid\topins\else\unvbox\topins\vskip\skip\topins\fi
   \dimen@ = \dp255 \unvbox255
   \vskip\pagebottomfiller
   \ifvoid\footins\else\vskip\skip\footins\footrule\unvbox\footins\fi
   \ifr@ggedbottom \kern-\dimen@ \vfil \fi }
\def\makeheadline{\vbox to 0pt{ \skip@=\topskip
      \advance\skip@ by -12pt \advance\skip@ by -2\normalbaselineskip
      \vskip\skip@ \line{\vbox to 12pt{}\the\headline} \vss
      }\nointerlineskip}
\def\makefootline{\baselineskip = 1.5\normalbaselineskip
                 \line{\the\footline}}
\newif\iffrontpage
\newif\ifp@genum
\def\nopagenumbers{\p@genumfalse}
\def\pagenumbers{\p@genumtrue}
\pagenumbers
\newtoks\paperheadline
\newtoks\paperfootline
\newtoks\letterheadline
\newtoks\letterfootline
\newtoks\letterinfo
\newtoks\date
\paperheadline={\hfil}
\paperfootline={\hss\iffrontpage\else\ifp@genum\tenrm\folio\hss\fi\fi}
\letterheadline{\iffrontpage \hfil \else
    \rm \ifp@genum page~~\folio\fi \hfil\the\date \fi}
\letterfootline={\iffrontpage\the\letterinfo\else\hfil\fi}
\letterinfo={\hfil}
\def\monthname{\rel@x\ifcase\month 0/\or January\or February\or
   March\or April\or May\or June\or July\or August\or September\or
   October\or November\or December\else\number\month/\fi}
\def\today{\monthname~\number\day, \number\year}
\date={\today}
\headline=\paperheadline 
\footline=\paperfootline 
\countdef\pageno=1      \countdef\pagen@=0
\countdef\pagenumber=1  \pagenumber=1
\def\advancepageno{\gl@bal\advance\pagen@ by 1
   \ifnum\pagenumber<0 \gl@bal\advance\pagenumber by -1
    \else\gl@bal\advance\pagenumber by 1 \fi
    \gl@bal\frontpagefalse  \swing@ }
\def\folio{\ifnum\pagenumber<0 \romannumeral-\pagenumber
           \else \number\pagenumber \fi }
\def\swing@{\ifodd\pagenumber \gl@bal\advance\hoffset by -\HSWING
             \else \gl@bal\advance\hoffset by \HSWING \fi }
\def\footrule{\dimen@=\prevdepth\nointerlineskip
   \vbox to 0pt{\vskip -0.25\baselineskip \hrule width 0.35\hsize \vss}
   \prevdepth=\dimen@ }
\let\footnotespecial=\rel@x
\newdimen\footindent
\footindent=24pt
\def\Textindent#1{\noindent\llap{#1\enspace}\ignorespaces}
\def\Vfootnote#1{\insert\footins\bgroup
   \interlinepenalty=\interfootnotelinepenalty \floatingpenalty=20000
   \singl@true\doubl@false\Tenpoint
   \splittopskip=\ht\strutbox \boxmaxdepth=\dp\strutbox
   \leftskip=\footindent \rightskip=\z@skip
   \parindent=0.5\footindent \parfillskip=0pt plus 1fil
   \spaceskip=\z@skip \xspaceskip=\z@skip \footnotespecial
   \Textindent{#1}\footstrut\futurelet\next\fo@t}

\def\vfootnote#1{\Vfootnote{${#1}$}}
\def\footnote#1{\attach{#1}\vfootnote{#1}}

\let\footsymbol=\star
\newcount\lastf@@t           \lastf@@t=-1
\newcount\footsymbolcount    \footsymbolcount=0
\newif\ifPhysRev
\def\bumpfootsymbolcount{\rel@x
   \iffrontpage \bumpfootsymbolpos \else \advance\lastf@@t by 1
     \ifPhysRev \bumpfootsymbolneg \else \bumpfootsymbolpos \fi \fi
   \gl@bal\lastf@@t=\pagen@ }
\def\bumpfootsymbolpos{\ifnum\footsymbolcount <0
                            \gl@bal\footsymbolcount =0 \fi
    \ifnum\lastf@@t<\pagen@ \gl@bal\footsymbolcount=0
     \else \gl@bal\advance\footsymbolcount by 1 \fi }
\def\bumpfootsymbolneg{\ifnum\footsymbolcount >0
             \gl@bal\footsymbolcount =0 \fi
         \gl@bal\advance\footsymbolcount by -1 }
\def\fd@f#1 {\xdef\footsymbol{\mathchar"#1 }}
\def\generatefootsymbol{\ifcase\footsymbolcount \fd@f 13F \or \fd@f 279
        \or \fd@f 27A \or \fd@f 278 \or \fd@f 27B \else
        \ifnum\footsymbolcount <0 \fd@f{023 \number-\footsymbolcount }
         \else \fd@f 203 {\loop \ifnum\footsymbolcount >5
                \fd@f{203 \footsymbol } \advance\footsymbolcount by -1
                \repeat }\fi \fi }

\def\nonfrenchspacing{\sfcode`\.=3001 \sfcode`\!=3000 \sfcode`\?=3000
        \sfcode`\:=2000 \sfcode`\;=1500 \sfcode`\,=1251 }
\nonfrenchspacing
\newdimen\d@twidth
{\setbox0=\hbox{s.} \gl@bal\d@twidth=\wd0 \setbox0=\hbox{s}
        \gl@bal\advance\d@twidth by -\wd0 }
\def\removehglue{\loop \unskip \ifdim\lastskip >\z@ \repeat }
\def\roll@ver#1{\removehglue \nobreak \count255 =\spacefactor \dimen@=\z@
        \ifnum\count255 =3001 \dimen@=\d@twidth \fi
        \ifnum\count255 =1251 \dimen@=\d@twidth \fi
    \iftwelv@ \kern-\dimen@ \else \kern-0.83\dimen@ \fi
   #1\spacefactor=\count255 }
\def\step@ver#1{\rel@x \ifmmode #1\else \ifhmode
        \roll@ver{${}#1$}\else {\setbox0=\hbox{${}#1$}}\fi\fi }
\def\attach#1{\step@ver{\strut^{\mkern 2mu #1} }}
%
%
%
\newcount\chapternumber      \chapternumber=0
\newcount\sectionnumber      \sectionnumber=0
\newcount\equanumber         \equanumber=0
\let\chapterlabel=\rel@x
\let\sectionlabel=\rel@x
\newtoks\chapterstyle        \chapterstyle={\Number}
\newtoks\sectionstyle        \sectionstyle={\chapterlabel.\Number}
\newskip\chapterskip         \chapterskip=\bigskipamount
\newskip\sectionskip         \sectionskip=\medskipamount
\newskip\headskip            \headskip=8pt plus 3pt minus 3pt
\newdimen\chapterminspace    \chapterminspace=15pc
\newdimen\sectionminspace    \sectionminspace=10pc
\newdimen\referenceminspace  \referenceminspace=20pc
\def\chapterreset{\gl@bal\advance\chapternumber by 1
   \ifnum\equanumber<0 \else\gl@bal\equanumber=0\fi
   \sectionnumber=0 \let\sectionlabel=\rel@x
   {\pr@tect\xdef\chapterlabel{\the\chapterstyle{\the\chapternumber}}}}
\def\alphabetic#1{\count255='140 \advance\count255 by #1\char\count255}
\def\Alphabetic#1{\count255='100 \advance\count255 by #1\char\count255}
\def\Roman#1{\uppercase\expandafter{\romannumeral #1}}
\def\roman#1{\romannumeral #1}
\def\Number#1{\number #1}
\def\BLANC#1{}
\def\titleparagraphs{\interlinepenalty=9999
     \leftskip=0.03\hsize plus 0.22\hsize minus 0.03\hsize
     \rightskip=\leftskip \parfillskip=0pt
     \hyphenpenalty=9000 \exhyphenpenalty=9000
     \tolerance=9999 \pretolerance=9000
     \spaceskip=0.333em \xspaceskip=0.5em }
\def\titlestyle#1{\par\begingroup \titleparagraphs
     \iftwelv@\fourteenpoint\else\twelvepoint\fi
   \noindent #1\par\endgroup }
\def\spacecheck#1{\dimen@=\pagegoal\advance\dimen@ by -\pagetotal
   \ifdim\dimen@<#1 \ifdim\dimen@>0pt \vfil\break \fi\fi}
\def\chapter#1{\par \penalty-300 \vskip\chapterskip
   \spacecheck\chapterminspace
   \chapterreset \titlestyle{\chapterlabel.~#1}
   \nobreak\vskip\headskip \penalty 30000
   {\pr@tect\wlog{\string\chapter\space \chapterlabel}} }

\def\section#1{\par \ifnum\the\lastpenalty=30000\else
   \penalty-200\vskip\sectionskip \spacecheck\sectionminspace\fi
   \gl@bal\advance\sectionnumber by 1
   {\pr@tect
   \xdef\sectionlabel{\the\sectionstyle\the\sectionnumber}
   \wlog{\string\section\space \sectionlabel}}
   \noindent {\caps\enspace\sectionlabel.~~#1}\par
   \nobreak\vskip\headskip \penalty 30000 }
\def\subsection#1{\par
   \ifnum\the\lastpenalty=30000\else \penalty-100\smallskip \fi
   \noindent\undertext{#1}\enspace \vadjust{\penalty5000}}

\def\undertext#1{\vtop{\hbox{#1}\kern 1pt \hrule}}
\def\APPENDIX#1#2{\par\penalty-300\vskip\chapterskip
   \spacecheck\chapterminspace \chapterreset \xdef\chapterlabel{#1}
   \titlestyle{APPENDIX #2} \nobreak\vskip\headskip \penalty 30000
   \wlog{\string\Appendix~\chapterlabel} }
\def\Appendix#1{\APPENDIX{#1}{#1}}
\def\appendix{\APPENDIX{A}{}}
\def\unnumberedchapters{\let\makechapterlabel=\rel@x
      \let\chapterlabel=\rel@x  \sectionstyle={\BLANC}
      \let\sectionlabel=\rel@x \sequentialequations }
%
%
%
\def\eqname#1{\rel@x {\pr@tect
  \ifnum\equanumber<0 \xdef#1{{\rm(\number-\equanumber)}}%
     \gl@bal\advance\equanumber by -1
  \else \gl@bal\advance\equanumber by 1
     \ifx\chapterlabel\rel@x \def\d@t{}\else \def\d@t{.}\fi
    \xdef#1{{\rm(\chapterlabel\d@t\number\equanumber)}}\fi #1}}

\def\eqn{\eqno\eqname}

\def\eqinsert#1{\noalign{\dimen@=\prevdepth \nointerlineskip
   \setbox0=\hbox to\displaywidth{\hfil #1}
   \vbox to 0pt{\kern 0.5\baselineskip\hbox{$\!\box0\!$}\vss}
   \prevdepth=\dimen@}}
%

%
%
\def\GENITEM#1;#2{\par \hangafter=0 \hangindent=#1
    \Textindent{$ #2 $}\ignorespaces}
\outer\def\newitem#1=#2;{\gdef#1{\GENITEM #2;}}

\newdimen\itemsize                \itemsize=30pt
\newitem\item=1\itemsize;
\newitem\sitem=1.75\itemsize;     
\newitem\ssitem=2.5\itemsize;     
\outer\def\newlist#1=#2&#3&#4;{\toks0={#2}\toks1={#3}%
   \count255=\escapechar \escapechar=-1
   \alloc@0\list\countdef\insc@unt\listcount     \listcount=0
   \edef#1{\par
      \countdef\listcount=\the\allocationnumber
      \advance\listcount by 1
      \hangafter=0 \hangindent=#4
      \Textindent{\the\toks0{\listcount}\the\toks1}}
   \expandafter\expandafter\expandafter
    \edef\c@t#1{begin}{\par
      \countdef\listcount=\the\allocationnumber \listcount=1
      \hangafter=0 \hangindent=#4
      \Textindent{\the\toks0{\listcount}\the\toks1}}
   \expandafter\expandafter\expandafter
    \edef\c@t#1{con}{\par \hangafter=0 \hangindent=#4 \noindent}
   \escapechar=\count255}
\def\c@t#1#2{\csname\string#1#2\endcsname}
\newlist\point=\Number&.&1.0\itemsize;
\newlist\subpoint=(\alphabetic&)&1.75\itemsize;
\newlist\subsubpoint=(\roman&)&2.5\itemsize;
%

%
%
%
%
\newcount\referencecount     \referencecount=0
\newcount\lastrefsbegincount \lastrefsbegincount=0
\newif\ifreferenceopen       \newwrite\referencewrite
\newdimen\refindent          \refindent=30pt
\def\normalrefmark#1{\attach{\scriptscriptstyle [ #1 ] }}
\let\PRrefmark=\attach
\def\NPrefmark#1{\step@ver{{\;[#1]}}}
\def\refmark#1{\rel@x\ifPhysRev\PRrefmark{#1}\else\normalrefmark{#1}\fi}
\def\refend@{\refmark{\number\referencecount}}
\def\refend{\refend@{}\space }
\def\refsend{\refmark{\count255=\referencecount
   \advance\count255 by-\lastrefsbegincount
   \ifcase\count255 \number\referencecount
   \or \number\lastrefsbegincount,\number\referencecount
   \else \number\lastrefsbegincount-\number\referencecount \fi}\space }
\def\REFNUM#1{\rel@x \gl@bal\advance\referencecount by 1
    \xdef#1{\the\referencecount }}
\def\Refnum#1{\REFNUM #1\refend@ } 
\def\REF#1{\REFNUM #1\R@FWRITE\ignorespaces}
\def\Ref#1{\Refnum #1\REFWRITE }
\def\ref{\Ref\?}
\def\REFS#1{\REFNUM #1\gl@bal\lastrefsbegincount=\referencecount
    \REFWRITE }

\def\r@fitem#1{\par \hangafter=0 \hangindent=\refindent \Textindent{#1}}
\def\refitem#1{\r@fitem{#1.}}
\def\NPrefitem#1{\r@fitem{[#1]}}
\def\NPrefs{\let\refmark=\NPrefmark \let\refitem=\NPrefitem}
\def\REFWRITE{\R@FWRITE\rel@x }
\def\R@FWRITE#1{\ifreferenceopen \else \gl@bal\referenceopentrue
     \immediate\openout\referencewrite=\jobname.refs
     \toks@={\begingroup \refoutspecials \catcode`\^^M=10 }%
     \immediate\write\referencewrite{\the\toks@}\fi
    \immediate\write\referencewrite{\noexpand\refitem %
                                    {\the\referencecount}}%
    \p@rse@ndwrite \referencewrite #1}
\begingroup
 \catcode`\^^M=\active \let^^M=\relax %
 \gdef\p@rse@ndwrite#1#2{\begingroup \catcode`\^^M=12 \newlinechar=`\^^M%
         \chardef\rw@write=#1\sc@nlines#2}%
 \gdef\sc@nlines#1#2{\sc@n@line \g@rbage #2^^M\endsc@n \endgroup #1}%
 \gdef\sc@n@line#1^^M{\expandafter\toks@\expandafter{\deg@rbage #1}%
         \immediate\write\rw@write{\the\toks@}%
         \futurelet\n@xt \sc@ntest }%
\endgroup
\def\sc@ntest{\ifx\n@xt\endsc@n \let\n@xt=\rel@x
       \else \let\n@xt=\sc@n@notherline \fi \n@xt }
\def\sc@n@notherline{\sc@n@line \g@rbage }
\def\deg@rbage#1{}
\let\g@rbage=\relax    \let\endsc@n=\relax
\def\refout{\par\penalty-400\vskip\chapterskip
   \spacecheck\referenceminspace
   \ifreferenceopen \Closeout\referencewrite \referenceopenfalse \fi
   \line{\fourteenrm\hfil REFERENCES\hfil}\vskip\headskip
   \input \jobname.refs
   }
\def\refoutspecials{\sfcode`\.=1000 \interlinepenalty=1000
         \rightskip=\z@ plus 1em minus \z@ }
\def\Closeout#1{\toks0={\par\endgroup}\immediate\write#1{\the\toks0}%
   \immediate\closeout#1}
%
%
\newcount\figurecount     \figurecount=0
\newcount\tablecount      \tablecount=0
\newif\iffigureopen       \newwrite\figurewrite
\newif\iftableopen        \newwrite\tablewrite
\def\FIGNUM#1{\rel@x \gl@bal\advance\figurecount by 1
    \xdef#1{\the\figurecount}}
\def\FIGURE#1{\FIGNUM #1\F@GWRITE\ignorespaces }

\def\figitem#1{\r@fitem{#1)}}
\def\FIGWRITE{\F@GWRITE\rel@x }
\def\TABNUM#1{\rel@x \gl@bal\advance\tablecount by 1
    \xdef#1{\the\tablecount}}
\def\TABLE#1{\TABNUM #1\T@BWRITE\ignorespaces }

\def\tabitem#1{\r@fitem{#1:}}
\def\TABWRITE{\T@BWRITE\rel@x }
\def\F@GWRITE#1{\iffigureopen \else \gl@bal\figureopentrue
     \immediate\openout\figurewrite=\jobname.figs
     \toks@={\begingroup \catcode`\^^M=10 }%
     \immediate\write\figurewrite{\the\toks@}\fi
    \immediate\write\figurewrite{\noexpand\figitem %
                                 {\the\figurecount}}%
    \p@rse@ndwrite \figurewrite #1}
\def\T@BWRITE#1{\iftableopen \else \gl@bal\tableopentrue
     \immediate\openout\tablewrite=\jobname.tabs
     \toks@={\begingroup \catcode`\^^M=10 }%
     \immediate\write\tablewrite{\the\toks@}\fi
    \immediate\write\tablewrite{\noexpand\tabitem %
                                 {\the\tablecount}}%
    \p@rse@ndwrite \tablewrite #1}
\def\figout{\par\penalty-400
   \vskip\chapterskip\spacecheck\referenceminspace
   \iffigureopen \Closeout\figurewrite \figureopenfalse \fi
   \line{\fourteenrm\hfil FIGURE CAPTIONS\hfil}\vskip\headskip
   \input \jobname.figs
   }
\def\tabout{\par\penalty-400
   \vskip\chapterskip\spacecheck\referenceminspace
   \iftableopen \Closeout\tablewrite \tableopenfalse \fi
   \line{\fourteenrm\hfil TABLE CAPTIONS\hfil}\vskip\headskip
   \input \jobname.tabs
   }
%
%
%
\newbox\picturebox
\def\p@cht{\ht\picturebox }
\def\p@cwd{\wd\picturebox }
\def\p@cdp{\dp\picturebox }
\newdimen\xshift
\newdimen\yshift
\newdimen\captionwidth
\newskip\captionskip
\captionskip=15pt plus 5pt minus 3pt
\def\fullwidth{\captionwidth=\hsize }
\newtoks\Caption
\newif\ifcaptioned
\newif\ifselfcaptioned
\def\caption{\captionedtrue \Caption }
\newcount\linesabove
\newif\iffileexists
\newtoks\picfilename
\def\fil@#1 {\fileexiststrue \picfilename={#1}}
\def\file#1{\if=#1\let\n@xt=\fil@ \else \def\n@xt{\fil@ #1}\fi \n@xt }
\def\pl@t{\begingroup \pr@tect
    \setbox\picturebox=\hbox{}\fileexistsfalse
    \let\height=\p@cht \let\width=\p@cwd \let\depth=\p@cdp
    \xshift=\z@ \yshift=\z@ \captionwidth=\z@
    \Caption={}\captionedfalse
    \linesabove =0 \picturedefault }
\def\plot{\pl@t \selfcaptionedfalse }
\def\Picture#1{\gl@bal\advance\figurecount by 1
    \xdef#1{\the\figurecount}\pl@t \selfcaptionedtrue }

\def\s@vepicture{\iffileexists \parsefilename \redopicturebox \fi
   \ifdim\captionwidth>\z@ \else \captionwidth=\p@cwd \fi
   \xdef\lastpicture{\iffileexists
        \setbox0=\hbox{\raise\the\yshift \vbox{%
              \moveright\the\xshift\hbox{\picturedefinition}}}%
        \else \setbox0=\hbox{}\fi
         \ht0=\the\p@cht \wd0=\the\p@cwd \dp0=\the\p@cdp
         \vbox{\hsize=\the\captionwidth \line{\hss\box0 \hss }%
              \ifcaptioned \vskip\the\captionskip \noexpand\Tenpoint
                \ifselfcaptioned Figure~\the\figurecount.\enspace \fi
                \the\Caption \fi }}%
    \endgroup }
\let\endpicture=\s@vepicture
\def\savepicture#1{\s@vepicture \global\let#1=\lastpicture }
\def\displaypicture{\fullwidth \s@vepicture $$\lastpicture $${}}
\def\toppicture{\fullwidth \s@vepicture \topinsert
    \lastpicture \medskip \endinsert }
\def\midpicture{\fullwidth \s@vepicture \midinsert
    \lastpicture \endinsert }
%
%
\def\leftpicture{\pres@tpicture
    \dimen@i=\hsize \advance\dimen@i by -\dimen@ii
    \setbox\picturebox=\hbox to \hsize {\box0 \hss }%
    \wr@paround }
\def\rightpicture{\pres@tpicture
    \dimen@i=\z@
    \setbox\picturebox=\hbox to \hsize {\hss \box0 }%
    \wr@paround }
\def\pres@tpicture{\gl@bal\linesabove=\linesabove
    \s@vepicture \setbox\picturebox=\vbox{
         \kern \linesabove\baselineskip \kern 0.3\baselineskip
         \lastpicture \kern 0.3\baselineskip }%
    \dimen@=\p@cht \dimen@i=\dimen@
    \advance\dimen@i by \pagetotal
    \par \ifdim\dimen@i>\pagegoal \vfil\break \fi
    \dimen@ii=\hsize
    \advance\dimen@ii by -\parindent \advance\dimen@ii by -\p@cwd
    \setbox0=\vbox to\z@{\kern-\baselineskip \unvbox\picturebox \vss }}
\def\wr@paround{\Caption={}\count255=1
    \loop \ifnum \linesabove >0
         \advance\linesabove by -1 \advance\count255 by 1
         \advance\dimen@ by -\baselineskip
         \expandafter\Caption \expandafter{\the\Caption \z@ \hsize }%
      \repeat
    \loop \ifdim \dimen@ >\z@
         \advance\count255 by 1 \advance\dimen@ by -\baselineskip
         \expandafter\Caption \expandafter{%
             \the\Caption \dimen@i \dimen@ii }%
      \repeat
    \edef\n@xt{\parshape=\the\count255 \the\Caption \z@ \hsize }%
    \par\noindent \n@xt \strut \vadjust{\box\picturebox }}
\let\picturedefault=\relax
\let\parsefilename=\relax
\def\redopicturebox{\let\picturedefinition=\rel@x
   \errhelp=\disabledpictures
   \errmessage{This version of TeX cannot handle pictures.  Sorry.}}
\newhelp\disabledpictures
     {You will get a blank box in place of your picture.}
%
%
%
%
%
%
%
%
%
%
\def\FRONTPAGE{\ifvoid255\else\vfill\penalty-20000\fi
   \gl@bal\pagenumber=1     \gl@bal\chapternumber=0
   \gl@bal\equanumber=0     \gl@bal\sectionnumber=0
   \gl@bal\referencecount=0 \gl@bal\figurecount=0
   \gl@bal\tablecount=0     \gl@bal\frontpagetrue
   \gl@bal\lastf@@t=0       \gl@bal\footsymbolcount=0}

\def\papers{\papersize\headline=\paperheadline\footline=\paperfootline}
\def\papersize{
   \advance\hoffset by\HOFFSET \advance\voffset by\VOFFSET
   \pagebottomfiller=0pc
   \skip\footins=\bigskipamount \normalspace }
\papers  
%
%
\newskip\lettertopskip       \lettertopskip=20pt plus 50pt
\newskip\letterbottomskip    \letterbottomskip=\z@ plus 100pt
\newskip\signatureskip       \signatureskip=40pt plus 3pt
\def\lettersize{\hsize=6.5in \vsize=8.5in \hoffset=0in \voffset=0.5in
   \advance\hoffset by\HOFFSET \advance\voffset by\VOFFSET
   \pagebottomfiller=\letterbottomskip
   \skip\footins=\smallskipamount \multiply\skip\footins by 3
   \singlespace }
\def\MEMO{\lettersize \headline=\letterheadline \footline={\hfil }%
   \let\rule=\memorule \FRONTPAGE \memohead }

\def\memodate{\afterassignment\MEMO \date }
\def\memit@m#1{\smallskip \hangafter=0 \hangindent=1in
    \Textindent{\caps #1}}
\def\subject{\memit@m{Subject:}}
\def\topic{\memit@m{Topic:}}
\def\from{\memit@m{From:}}
\def\memorule{\medskip\hrule height 1pt\bigskip}  
\def\memohead{\centerline{\fourteenrm MEMORANDUM}}
\newwrite\labelswrite
\newtoks\rw@toks
\def\letters{\lettersize
   \headline=\letterheadline \footline=\letterfootline
   \immediate\openout\labelswrite=\jobname.lab}

\let\letterhead=\rel@x
\def\addressee#1{\medskip\line{\hskip 0.75\hsize plus\z@ minus 0.25\hsize
                               \the\date \hfil }%
   \vskip \lettertopskip
   \ialign to\hsize{\strut ##\hfil\tabskip 0pt plus \hsize \crcr #1\crcr}
   \writelabel{#1}\medskip \noindent\hskip -\spaceskip \ignorespaces }
\def\rwl@begin#1\cr{\rw@toks={#1\crcr}\rel@x
   \immediate\write\labelswrite{\the\rw@toks}\futurelet\n@xt\rwl@next}
\def\rwl@next{\ifx\n@xt\rwl@end \let\n@xt=\rel@x
      \else \let\n@xt=\rwl@begin \fi \n@xt}
\let\rwl@end=\rel@x
\def\writelabel#1{\immediate\write\labelswrite{\noexpand\labelbegin}
     \rwl@begin #1\cr\rwl@end
     \immediate\write\labelswrite{\noexpand\labelend}}
\newtoks\FromAddress         \FromAddress={}
\newtoks\sendername          \sendername={}
\newbox\FromLabelBox
\newdimen\labelwidth          \labelwidth=6in
\def\makelabels{\afterassignment\Makelabels \sendersname=}
\def\Makelabels{\FRONTPAGE \letterinfo={\hfil } \MakeFromBox
     \immediate\closeout\labelswrite  \input \jobname.lab\vfil\eject}
\let\labelend=\rel@x
\def\labelbegin#1\labelend{\setbox0=\vbox{\ialign{##\hfil\cr #1\crcr}}
     \MakeALabel }
\def\MakeFromBox{\gl@bal\setbox\FromLabelBox=\vbox{\Tenpoint
     \ialign{##\hfil\cr \the\sendername \the\FromAddress \crcr }}}
\def\MakeALabel{\vskip 1pt \hbox{\vrule \vbox{
        \hsize=\labelwidth \hrule\bigskip
        \leftline{\hskip 1\parindent \copy\FromLabelBox}\bigskip
        \centerline{\hfil \box0 } \bigskip \hrule
        }\vrule } \vskip 1pt plus 1fil }
\def\signed#1{\par \nobreak \bigskip \dt@pfalse \begingroup
  \everycr={\noalign{\nobreak
            \ifdt@p\vskip\signatureskip\gl@bal\dt@pfalse\fi }}%
  \tabskip=0.5\hsize plus \z@ minus 0.5\hsize
  \halign to\hsize {\strut ##\hfil\tabskip=\z@ plus 1fil minus \z@\crcr
          \noalign{\gl@bal\dt@ptrue}#1\crcr }%
  \endgroup \bigskip }
\newbox\letterb@x
\def\lettertext{\par \vskip\parskip \unvcopy\letterb@x \par }
\def\multiletter{\setbox\letterb@x=\vbox\bgroup
      \everypar{\vrule height 1\baselineskip depth 0pt width 0pt }
      \singlespace \topskip=\baselineskip }
\def\letterend{\par\egroup}
%
%
%
\newskip\frontpageskip
\newtoks\Pubnum   
\newtoks\Pubtype  \let\pubtype=\Pubtype
\newif\ifp@bblock  \p@bblocktrue
\def\PH@SR@V{\doubl@true \baselineskip=24.1pt plus 0.2pt minus 0.1pt
             \parskip= 3pt plus 2pt minus 1pt }
\def\PHYSREV{\papers\PhysRevtrue\PH@SR@V}

\def\titlepage{\FRONTPAGE\papers\ifPhysRev\PH@SR@V\fi
   \ifp@bblock\p@bblock \else\hrule height\z@ \rel@x \fi }
\def\nopubblock{\p@bblockfalse}

\frontpageskip=12pt plus .5fil minus 2pt
\Pubtype={}
\Pubnum={}
\def\p@bblock{\begingroup \tabskip=\hsize minus \hsize
   \baselineskip=1.5\ht\strutbox \topspace-2\baselineskip
   \halign to\hsize{\strut ##\hfil\tabskip=0pt\crcr
       \the\Pubnum\crcr\the\date\crcr\the\pubtype\crcr}\endgroup}
\def\title#1{\vskip\frontpageskip \titlestyle{#1} \vskip\headskip }
\def\author#1{\vskip\frontpageskip\titlestyle{\twelvecp #1}\nobreak}

\def\address#1{\par\kern 5pt\titlestyle{\twelvepoint\it #1}}
\def\andaddress{\par\kern 5pt \centerline{\sl and} \address}

\def\abstract{\par\dimen@=\prevdepth \hrule height\z@ \prevdepth=\dimen@
   \vskip\frontpageskip\centerline{\fourteenrm ABSTRACT}\vskip\headskip }

%
%
%

\def\\{\rel@x \ifmmode \backslash \else {\tt\char`\\}\fi }
\def\sequentialequations{\rel@x \if\equanumber<0 \else
  \gl@bal\equanumber=-\equanumber \gl@bal\advance\equanumber by -1 \fi }
\def\journal#1&#2(#3){\begingroup \let\journal=\dummyj@urnal
    \unskip, \sl #1\unskip~\bf\ignorespaces #2\rm
    (\afterassignment\j@ur \count255=#3), \endgroup\ignorespaces }
\def\j@ur{\ifnum\count255<100 \advance\count255 by 1900 \fi
          \number\count255 }
\def\dummyj@urnal{%
    \toks@={Reference foul up: nested \journal macros}%
    \errhelp={Your forgot & or ( ) after the last \journal}%
    \errmessage{\the\toks@ }}

\def\topspace{\hrule height 0pt depth 0pt \vskip}

\def\Buildrel#1\under#2{\mathrel{\mathop{#2}\limits_{#1}}}
\def\becomes#1{\mathchoice{\becomes@\scriptstyle{#1}}
   {\becomes@\scriptstyle{#1}} {\becomes@\scriptscriptstyle{#1}}
   {\becomes@\scriptscriptstyle{#1}}}
\def\becomes@#1#2{\mathrel{\setbox0=\hbox{$\m@th #1{\,#2\,}$}%
        \mathop{\hbox to \wd0 {\rightarrowfill}}\limits_{#2}}}

\let\int=\intop         
\def\lsim{\mathrel{\mathpalette\@versim<}}
\def\gsim{\mathrel{\mathpalette\@versim>}}
\def\@versim#1#2{\vcenter{\offinterlineskip
        \ialign{$\m@th#1\hfil##\hfil$\crcr#2\crcr\sim\crcr } }}
\def\big#1{{\hbox{$\left#1\vbox to 0.85\b@gheight{}\right.\n@space$}}}
\def\Big#1{{\hbox{$\left#1\vbox to 1.15\b@gheight{}\right.\n@space$}}}
\def\bigg#1{{\hbox{$\left#1\vbox to 1.45\b@gheight{}\right.\n@space$}}}
\def\Bigg#1{{\hbox{$\left#1\vbox to 1.75\b@gheight{}\right.\n@space$}}}
\def\){\mskip 2mu\nobreak }
%
%
%
\let\sec@nt=\sec
\def\sec{\rel@x\ifmmode\let\n@xt=\sec@nt\else\let\n@xt\section\fi\n@xt}
\def\obsolete#1{\message{Macro \string #1 is obsolete.}}
\def\firstsec#1{\obsolete\firstsec \section{#1}}
\def\firstsubsec#1{\obsolete\firstsubsec \subsection{#1}}
\def\thispage#1{\obsolete\thispage \gl@bal\pagenumber=#1\frontpagefalse}
\def\thischapter#1{\obsolete\thischapter \gl@bal\chapternumber=#1}
\def\splitout{\obsolete\splitout\rel@x}
\def\prop{\obsolete\prop \propto }
\def\nextequation#1{\obsolete\nextequation \gl@bal\equanumber=#1
   \ifnum\the\equanumber>0 \gl@bal\advance\equanumber by 1 \fi}
\def\BOXITEM{\afterassigment\B@XITEM\setbox0=}
\def\B@XITEM{\par\hangindent\wd0 \noindent\box0 }
%
%
%
\def\phyzzx{PHY\setbox0=\hbox{Z}\copy0 \kern-0.5\wd0 \box0 X}
        
\everyjob{\xdef\today{\monthname~\number\day, \number\year}
        \input myphyx.tex }
\message{ by V.K.}
%
\catcode`\@=12 
%
\doublespace

\centerline{\bf EMPIRICAL CONSTRAINTS ON COSMOLOGICAL GAMMA-RAY BURSTS}
\bigskip
\bigskip
\centerline{Eric Woods and Abraham Loeb}
\medskip
\centerline{Astronomy Department, Harvard University, 60 Garden St., Cambridge
MA 02138}
\bigskip
\centerline{\bf ABSTRACT}

We place empirical constraints on the physical properties
of $\gamma-$ray burst events at cosmological distances.
In particular we derive probability distributions for
the radiation energy $E_\gamma$, the minimum Lorentz factor $\gamma_{\rm min}$,
the maximum baryonic mass $M_{\rm max}$ and the
upper bound on the surrounding gas density $n_{\rm max}$
in the events, based on 169
bursts from the first BATSE catalog. Using peak flux as a distance indicator
we probe bursts where the constraints are stronger than average.
The resulting variance and skewness of the cosmological
probability distributions are calculated in addition to their mean values:
$\langle E_\gamma\rangle=4\times 10^{51} h^{-2} {\rm erg}$,
$\langle \gamma_{\rm min}\rangle= 5\times 10^2$,
$\langle M_{\rm max}\rangle=10^{-5}\xi^{-1} M_\odot$,
and $\langle (\gamma/\gamma_{\rm min})^{-5} n_{\rm max}\rangle\lsim
10^{2} \xi^{-1}{\rm cm^{-3}}$,
where $\xi$ is the fraction of the total energy
which is converted to $\gamma-$rays.
The distribution of burst energies ends at about $10^{53} {\rm erg}$,
close to the binding energy of a neutron star.

\bigskip
\noindent
{\it Subject headings:} cosmology: observations--gamma rays: bursts
\bigskip
\bigskip
\centerline{Submitted to \it The Astrophysical Journal Letters}
\np

\centerline{\bf 1. INTRODUCTION}

The origin of $\gamma-$ray bursts has been a long-standing puzzle
over the past two decades. Recently,
the Burst and Transient Source Experiment (BATSE) on board the
Compton Gamma-Ray Observatory (GRO) has provided important clues
about the burst population.
The distribution of the BATSE bursts appears to be entirely isotropic
and shows a deficiency of
faint bursts relative to a uniform Euclidean
distribution (Meegan et al. 1992; see also Meegan et al. 1993).
Both of these facts are most naturally explained
by a cosmological distance
scale for which the deficiency of faint bursts results from
a cosmological redshift (Mao \&
Paczy\'nski 1992; Piran 1992). A cosmological redshift is also
implied by claims that dim bursts are on average longer than
bright bursts (Norris et al. 1993).

For a cosmological distance scale, the observed burst properties indicate
that the emitting region moves with a highly relativistic
speed toward the observer.
A stationary source of a size compatible with the burst
duration $T$ has an optical depth to its own photons
of $\sim 10^{10} (T/{\rm sec})^{-2}$. Photon-photon collisions through
$\gamma + \gamma \rightarrow e^+ + e^-$ would
create pairs and thermalize the emission spectrum (Goodman 1986; Paczy\'nski
1986),
in conflict with the observed non-thermal spectra ($\nu I_\nu\approx const$
between $0.1-100 {\rm MeV}$).
However, if the emitting region expands relativistically
toward the observer the photons are beamed
and their relative momentum can be lowered below
the pair production threshold.
The Lorentz factor of the emitting region must obey
$\gamma\gsim10^{2}$ in order that a typical $\gamma-$ray
source be optically thin to its own photons (Fenimore, Epstein, \& Ho
1993; M\'esz\'aros, Laguna, \& Rees 1994).
This requirement, in turn,
places an upper bound on the amount of baryonic mass released in the explosive
event (Shemi \& Piran 1990).
If the observed $\gamma-$ray energy $E_\gamma$ reflects a fraction $\xi$ of
the total energy released then the rest mass
of the relativistic matter must be $M= E_\gamma/\xi\gamma c^2$, where $c$
is the speed of light.
At a cosmological distance the typical $\gamma-$ray energy of
$\langle E_\gamma\rangle\approx 10^{51}{\rm erg}$, and $\gamma>10^2$,
translate to a maximum amount
of baryonic contamination $M\lsim0.6\times10^{-5}\xi^{-1} M_{\odot}$.
This limit can be used to infer an upper bound on the ambient gas
density near the burst source.

The above constraints on the total energy, the minimum Lorentz
factor, and the maximum amount
of baryonic contamination provide useful guidelines for theoretical
models concerning the origin of cosmological $\gamma-$ray bursts.
In the past, these constraints were only phrased in rough numbers
assuming typical values for the burst properties (Fenimore, Epstein,
\& Ho 1993).
However, by now the publicly available BATSE catalog
allows an accurate statistical evaluation of these constraints.
In this {\it Letter} we derive the distribution
of the above quantities for the entire burst population.
Using peak flux as a distance indicator,
we probe events where the constraints are
more severe than average.
Such events can potentially be in conflict with models
that have an absolute maximum to the available energy
or an unavoidable amount of baryonic contamination.
In \S 2 we express the minimum Lorentz factor, the maximum
baryonic mass, and the upper bound on the ambient gas density
as functions of observed burst properties,
such as peak flux and fluence.
We then calibrate the cosmological distance scale based
on a fit to the number-peak flux distribution of bursts
in a flat universe (\S 3).
Using this calibration we obtain
probability distributions for the above constraints,
and calculate
the variance and skewness of these
distributions
in addition to their mean.
Finally, \S 4 summarizes
the main conclusions of this work.

\bigskip
\centerline{\bf 2. OPTICAL THICKNESS OF COSMOLOGICAL BURSTS TO THEIR PHOTONS}

We first calculate
the optical depth for pair production by photon-photon
collisions near a cosmological source.
A test photon of an energy $\epsilon$ can only produce an $e^{+}e^{-}$
pair in a collision with a photon whose energy
$\epsilon^\prime$ is greater than $\epsilon_t\equiv 2m^2c^4/
\epsilon(1-\cos\theta)$, where $\theta$ is the
angle between the photon trajectories.  The cross
section for collisions above threshold is
$$
\sigma (\epsilon,\epsilon^\prime,\theta) = {3\over 16}\sigma_{_T}
(1-v^2)\left[(3-v^4)\ln{(1+v)\over(1-v)} - 2v(2-v^2)\right],
\eqn\xsec
$$
where $\sigma_{_T}$ is the Thomson cross section and
$v=[1-(\epsilon_t /\epsilon^\prime)]^{1/2}$ is the center-of-mass
speed of the outgoing pair particles in units of the speed
of light (e.g. Berestetskii, Lifshitz, \& Pitaevskii 1982).

Now consider a cosmological point source from which a large amount of
energy is released in the form of relativistically
expanding matter, such as baryons and $e^{+}e^{-}$ pairs.
We assume that the matter expands
with a Lorentz factor $\gamma$ and a speed $c\beta=c(1-1/\gamma^2)^{1/2}$
toward the observer,
and that at some distance $r_0$ away from the source the matter
produces $\gamma-$rays isotropically in its rest frame (e.g. due to its
interaction
with the interstellar medium; cf. M\'esz\'aros , Laguna, \& Rees 1994 and
references therein).
For $\gamma\gg1$ the relativistic beaming of this radiation
is high in the observer's frame, and most
of the observed $\gamma-$rays
originate from a small solid angle
around the observer's line of sight. Thus, complete spherical symmetry
is not mandatory and jets are described as well
by our discussion.
Let us now find the optical thickness
of the $\gamma-$ray burst to a test photon
that propagates precisely
along the line of sight and has an energy $\epsilon$.
We integrate the collision probability from a radius $r=r_0$
out to infinity,
$$
\tau(\epsilon) = \int_{r_0}^{\infty} dr \int_{\epsilon_t}^{\infty}
d\epsilon^\prime
\int d\Omega {{dn}\over{d\epsilon' d\Omega}} \sigma(\epsilon,\epsilon^\prime,
\theta),
\eqn\depthfirst
$$
where $dn/d\epsilon^\prime d\Omega$ is the number density of photons per unit
energy per unit solid angle as a function of radius. Most of
the contribution to the integral comes from the region
near the source.
For simplicity we assume that the source maintains a steady
luminosity of $\gamma-$rays over a period $>r_0/2\gamma^2c$.
Unfortunately it is not possible to test this assumption
empirically.
The observed duration of a
burst is longer than the geometric time delay
among photons with similar relativistic beaming.
At a distance $r$ the photons arrive from a cone with
an opening angle $\theta \lsim r_0/r\gamma$ relative to the line of sight,
and are therefore spread over a time interval $\gsim r_0/2\gamma^2c$.

The photon density per solid angle at a radius $r$ is just proportional
to the projected area of a narrow ring on the spherical photosphere
(centered around the line of sight) times the relativistic emission
probability,
$$
\eqalign{{{dn}\over{d\epsilon^\prime d\Omega}} = {{1}\over{2\pi}}&\left\{
{(r/r_0)^2\over \{1-2r_0[\beta+(1-\beta)\ln(1-\beta)]/\beta^2r\}}\left[{1\over
{\sqrt{1-(r\theta/r_0)^2}}}-{2r_0\over r}\right]\right\}\cr
/r_0)^2}]^2}}{{dn}\over{d\epsilon'}},\cr}
\eqn\dnde
$$
The factor in curly brackets is the normalized area of the emission ring
which subtends an angle $\theta$ at a distance $r$; we assume that $\theta\ll1$
due to relativistic beaming. The second factor
introduces this beaming effect through
a Lorentz transformation of the solid angle
from the rest frame of the emitting matter
to the observer's frame (Rybicki \& Lightman 1979).
The photon number density per unit energy may be expressed in
terms of the observed energy flux per unit energy $dF/d\epsilon_o$,
$$
{{dn}\over{d\epsilon'}} = {{1}\over{c\epsilon'}} {{D^2}\over{r^2}}
{{d\epsilon_o}\over{d\epsilon'}} {{dF}\over{d\epsilon_o}},
\eqn\flux
$$
where $\epsilon_o\equiv\epsilon^\prime/(1+z)$ is the observed energy of a
photon
which is emitted with an energy $\epsilon^\prime$ at a cosmological redshift
$z$,
and $D(z)$ is the luminosity distance of the source.
Equations $\depthfirst-\flux$ provide the total
optical depth of the burst to a test photon,
$$
\eqalign{\tau (\epsilon) = {D^2\over c}
\int_{r_0}^\infty {{dr}\over{r^2}} \int_0^1 udu &\left\{{{(1-u^2)^{-1/2}}-
{2r_0/ r}\over 1-2r_0[\beta+(1-\beta)\ln(1-\beta)]/\beta^2r}\right\}
{(1-\beta)\over{(1-\beta\sqrt{1-u^2})^2}} \cr
{{d\epsilon'}\over{\epsilon'(1+z)}} {dF\over {d\epsilon_o}}
\sigma({\epsilon,\epsilon',u}),\cr}
\eqn\depth
$$
where $u\equiv r\theta/r_{0}$ and
the burst spectrum
can usually be approximated
by a power law
$$
{dF\over d\epsilon_o} = k\epsilon_o^{\alpha},
{\ }{\ }{\ }{\ }{\ }{\ }{\ }{\ }{\ }{\ }{\ }{\ }{\ }
\epsilon_1 < \epsilon_o < \epsilon_2.
\eqn\spectrum
$$
The optical depth scales with the source redshift as
$\tau\propto (1+z)^{-(1+\alpha)}$.
Most burst spectra have roughly equal energy flux
per logarithmic energy interval, i.e.
$\alpha\approx -1$,
without a clear sign for a decline in the flux
at a particular $\epsilon_2$. For a test photon
energy $\epsilon\gg{\rm MeV}$
the value of $\tau$ is only weakly dependent
on the value of $\epsilon_2\gg 10{\rm MeV}$ since
$(dn/d\epsilon^\prime)\sigma\propto (\epsilon^\prime)^{-3}$. We therefore
take $\epsilon_2\rightarrow\infty$.

Equations $\depth$ and $\spectrum$ can be used to get
the minimum value of $r_0=r_{\rm min}(\gamma,\epsilon)$ for which the burst
satisfies $\tau<1$:
$$
 r_{\rm min} = {{11}\over{720}} {{k\epsilon}
\over{m_e^2c^4}} {{\sigma_{_T} D^2}\over{c}}
{1\over{\gamma^2}}\left[\ln(2\gamma^2)
-{27\over 8}\right],
\eqn\minvalue
$$
where $m_e$ is the electron mass, and we have
kept terms to leading order in $1/\gamma$.
Due to the geometric difference in travel times among photons
coming from all emission angles
between zero and $\sim 1/\gamma$, the observed duration of the $\gamma-$ray
burst satisfies $T\gsim (1+z)r_0/2c\gamma^2$, and therefore
$\gamma>\gamma_{\rm min}\equiv{\sqrt{(1+z)r_{\rm min}/2cT}}$. The
value of $\gamma_{\rm min}$ is then obtained using equation $\minvalue$.
Given a test photon energy, the minimum Lorentz factor $\gamma_{\rm min}$
which is consistent with $\tau<1$ is related to the burst duration $T$
and the source distance. To leading order,
$\gamma_{\rm min}\propto D^{1/2} T^{-1/4}$. Note that even an
abrupt variation
of the source luminosity would still be seen as a gradual
change in the burst flux over a finite time $\sim (1+z)r_0/2\gamma^2c$,
because most photons suffer a geometric time delay
as they do not move precisely along the line of sight. Therefore the time $T$
can be associated with the variability of the burst rather
than with its overall duration. In the following section we relate
$T$ to the period of peak flux emission in each burst.

The total energy of the relativistic matter is
$\gamma Mc^2$, where $M$ is its rest mass.
If a fraction $\xi$ of this energy is
converted into the observed $\gamma-$ray energy $E_\gamma$,
then the value of $r_{\rm min}$ translates into
a limit on the maximum baryonic mass allowed,
$M_{\rm max}=E_{\gamma}/(\xi\gamma_{\rm min} c^2)$.

A lower bound on the amount of baryonic matter is given by the
interstellar gas mass swept along by the expanding fireball,
$M_{_{\rm ISM}}=(4\pi/3)\overline{m} n_{_{\rm ISM}}r_{0}^3$, where
$\overline{m}$ is the mean atomic mass
and $n_{_{\rm ISM}}$ is the particle number
density. The requirement $M_{_{\rm ISM}}<M_{\rm max}$
yields an upper bound on the gas density surrounding the source:
$n_{_{\rm ISM}}<n_{\rm max}=
M/[4\pi \overline{m} r_{\rm min}^3/3]$,
out to $\langle r_0\rangle \lsim 10^{16}{\rm cm}$.
To leading order, $n_{\rm max}\propto \gamma^5$.

\np
\centerline{\bf 3. EMPIRICAL CONSTRAINTS}

The values of $\gamma_{\rm min}$, $M_{\rm max}$ and $n_{\rm max}$
can be calculated from equations $\depth$ and $\spectrum$
using the detected burst properties in the BATSE catalog
(Fishman et al. 1993).
In order to determine the luminosity distance $D$ to each burst,
an additional assumption needs to be made about
the distribution of bursts in redshift and luminosity.  The
data have been shown to
be consistent with a non-evolving population of
standard candles in peak flux,
all with identical power-law spectra (Mao \& Paczy\'nski 1992; Piran 1992).
For a universe with a zero cosmological constant, the statistical
significance of the number vs. flux fit is only weakly dependent on
the density parameter $\Omega$ (Wickramasinghe et al. 1993).
We therefore adopt the simplest model and assume a
non-evolving population of standard candles
with a spectral index $\alpha=-1$ in
an $\Omega =1$ universe.
In this case the observed peak number-flux of photons $C_{\rm peak}(z)$
from a source at a redshift $z$ is given by
$$
C_{\rm peak} = \int_{\epsilon_{_{L}}}^{\epsilon_{_U}}
{{d\epsilon_o}\over \epsilon_o}{dF\over d\epsilon_o}
={{\Gamma}\over{4\pi D^2/(1+z)}},
\eqn\peakflux
$$
where $\Gamma$ is the intrinsic peak
emission rate of photons per unit time of the source and $D$ is its
luminosity distance (Weinberg 1972)
$$
D = {{2c}\over{H_0}} \left[(1+z)-{\sqrt{1+z}}\right].
\eqn\lumidis
$$
We use  $C_{\rm peak}$ as a distance indicator rather than the ratio
of maximum to threshold count rates $C_{\rm max}/C_{\rm min}$,
since the values of $C_{\rm peak}$ were corrected for detector
orientation and atmospheric scattering.
The value $C^{64}_{\rm peak}$ in the 64 ms
channel of BATSE is taken as the best indicator
of the peak flux.  The uncertainty due to
trigger inefficiency becomes large for fluxes below
$\sim 1.0~{\rm cm^{-2} s^{-1}}$ (Fishman et al. 1993).  Eliminating
bursts with $C_{\rm peak}\equiv C^{64}_{\rm peak}<1.0~{\rm cm^{-2} s^{-1}}$
from the sample and performing a one-distribution Kolmogorov-Smirnov (K-S)
test gives a best-fit value of $\Gamma =
9\times 10^{56} h^{-2} {\rm s^{-1}}$
with a K-S significance of 70\%,
where $h\equiv H_0/$(100 km sec$^{-1}$ Mpc$^{-1}$) reflects
the Hubble constant. Figure 1 shows the fit obtained for the number count
vs. peak flux with this value of $\Gamma$.

The parameters in equation $\spectrum$ are taken
to be $\alpha=-1$ and
$k = \left({{1}/{\epsilon_{_{L}}}}-{{1}/{\epsilon_{_{U}}}}\right)^{-1}
C_{\rm peak}$, with $\epsilon_{_{L}}=0.05$ MeV and $\epsilon_{_{U}}=0.3$ MeV,
since the BATSE peak flux data is for the channels 50-300 keV.
To implement the results from \S 2 we also need an empirical limit on
the duration of peak flux emission,
defined as $T$.
An upper bound on this duration
can be obtained from the ratio between
the total burst fluence $S$ and the peak energy
flux,
$$
T<T_{\rm max} = \left({{1}\over{\epsilon_{_L}}}-{{1}
\over{\epsilon_{_U}}}\right){{1}
\over (1+z){\ln(\epsilon_{_U}/\epsilon_{_L})}}
{{S}\over{C_{\rm peak}}}.
\eqn\uppbound
$$
The resulting constraint on $\gamma_{\rm min}$ is an underestimate
since the actual period of
peak flux emission is shorter than $T_{\rm max}$
and $\gamma_{\rm min}\propto T^{-1/4}$.
The constraints we derive can therefore be improved by
using information on the temporal profiles of bursts beyond
the data available in the first BATSE catalog.

Equations $\depth-\uppbound$ were applied to the 169 bursts
in the first BATSE catalog with
$C_{\rm peak}>1 {\rm cm^{-2} s^{-1}}$. Figures 2b-d present the resulting
probability distributions (normalized to a unit area)
for the minimum Lorentz factor $\gamma_{\rm min}$, the
maximum baryonic mass $M_{\rm max}$, and the maximum ambient
density $n_{\rm max}$. We also show
the distribution of
the total $\gamma-$ray energy $E_{\gamma}$ (figure 2a).
The quoted values for $E_\gamma$ and $M_{\rm max}$ were derived
assuming spherical symmetry; our numbers should be multiplied
by $\Delta\Omega/4\pi$ for jets that cover a solid angle $\Delta\Omega$.
The plots were obtained for $h=0.75$ and a
test photon energy of $\epsilon/(1+z)=10$ MeV,
while to leading order
$E_{\gamma}\propto h^{-2}$, $\gamma_{\rm min}\propto h^{-1/2}\epsilon^{1/4}$,
$M_{\rm max}\propto h^{-3/2}\epsilon^{-1/4}$, and $n_{\rm max}\propto
h^{3/2}\epsilon^{-7/4}$.
Note that only
$n_{\rm max}$ is strongly
sensitive to the test photon energy $\epsilon$; the
values of $n_{\rm max}$
would be lowered appreciably if non-thermal spectra
actually extend to energies much higher than $10 {\rm MeV}$.
However, a modest increase in $\gamma/\gamma_{\rm min}$
can compensate for this reduction since $n_{\rm max}\propto \gamma^5$.

Table 1 gives the
mean $\langle x\rangle$, the standard
deviation $\sigma =\langle(x-\langle x\rangle)^2\rangle^{1/2}$,
and the skewness $s=\langle(x-\langle x\rangle)^3\rangle/\sigma^3$ for the
distributions shown in figure 2. A rough estimate of
the significance of the skewness is obtained by
comparing it to $(15/N_{tot})^{1/2}\approx 0.3$,
where $N_{tot}=169$ is the total number of data points (Press et al. 1992). The
distributions in $E_{\gamma}$, $M_{\rm max}$ and $n_{\rm max}$ are
significantly
skewed toward
high values, while $\gamma_{\rm min}$ is only mildly skewed.

\bigskip

\centerline{\bf 4. DISCUSSION}

We have studied the empirical constraints
from the first BATSE catalog on a
population of cosmological $\gamma-$ray burst events.
Our results were obtained under two simplifying assumptions.
First we used peak flux as a distance indicator.
If bursts are not
standard candles in peak flux, the conversion of $C_{\rm peak}$
to a source distance is not single valued and each burst
would be spread over more than one bin in the histograms of
figure 2.
The K-S significance of the
the $N-C_{\rm peak}$ fit in figure 1 does not decrease
sufficiently to rule out a luminosity function
with a width of the order of its mean.
The second assumption was that the peak flux emission of the source
is steady over a timescale $>r_0/2c\gamma^2$.
A shorter burst would have a smaller optical depth, because
photons emitted at an angle $\sim 1/\gamma$ would be unable to
catch up with the photons emitted along the line of sight.

Bearing these assumptions in mind, we obtain
a few physically interesting results.
The energy distribution in figure 2a
ends at about the binding
energy of a neutron star $\sim 10^{53}$ erg.
This result is consistent with a variety  of cosmological
models that associate $\gamma-$ray bursts
with neutron stars (Eichler et al. 1989; M\'esz\'aros \& Rees 1992; Narayan,
Paczy\'nski \& Piran 1992; Usov 1992; Loeb 1993), although in some
models it is difficult to extract all the binding
energy in $\gamma-$rays.
The minimum Lorentz factor distribution (figure 2c)
shows two peaks, which may be just a reflection
of the bimodal distribution of burst durations (Kouveliotou et al. 1993).
The distribution of ambient gas densities (figure 2d)
allows bursts to reside within the
interstellar medium of galaxies,
where the density
$n_{_{\rm ISM}}\approx 1~{\rm cm^{-3}}$.
Since $n_{\rm max}$ is only weakly dependent
on the fluence ($\propto S^{-1/4}$),
this distribution has a sharp cutoff
at its low end due to
the selection effect introduced
by taking bursts with $C_{\rm peak}>1~{\rm cm^{-2} s^{-1}}$.

The constraints presented in this {\it Letter} can
be improved significantly when more data
concerning temporal profiles and faint bursts
become available.
\np
\centerline{\bf REFERENCES}

\medskip
\noindent
Berestetskii, V. B., Lifshitz, E. M., \&
Pitaevskii, L. P. 1982, Quantum Electrodynamics (New York: Pergamon), p. 371
\medskip
\noindent
Eichler, D., Livio, M., Piran, T., \& Schramm, D. N. 1989, Nature, 340, 126
\medskip
\noindent
Fenimore, E., Epstein, R. I., \& Ho, C. 1993, Astron. Astrophys. Suppl., 97, 59
\medskip
\noindent
Fishman, G., et al. 1993, ApJ Supplement, in press
\medskip
\noindent
Goodman, J. 1986, ApJ, 308, L47
\medskip
\noindent
Loeb, A. 1993, Phys. Rev. D, 48, 3419
\medskip
\noindent
Kouveliotou, C., et al. 1993, in Proc. of the Huntsville Gamma-Ray Burst
Workshop, eds. G. Fishman, K. Hurley \& J. Brainerd (New York: AIP), in press
\medskip
\noindent
Mao, S., \& Paczy\'nski, B. 1992, ApJ, 388, L45
\medskip
\noindent
Meegan, C. A., et al. 1992, Nature, 355, 143
\medskip
\noindent
Meegan, C. A., et al. 1993, in Proc. of the Huntsville Gamma-Ray Burst
Workshop, eds. G. Fishman, K. Hurley \& J. Brainerd (New York: AIP), in press
\medskip
\noindent
M\'esz\'aros, P., Laguna, P., \& Rees, M. J. 1994, ApJ, in press
\medskip
\noindent
M\'esz\'aros, P., \& Rees, M. J. 1992, MNRAS, 257, 29p
\medskip
\noindent
Narayan, R., Paczy\'nski, B., \& Piran, T. 1992, ApJ, 395, L83
\medskip
\noindent
Norris, J., et al. 1993, in Proc. of the Huntsville Gamma-Ray Burst
Workshop, eds. G. Fishman, K. Hurley \& J. Brainerd (New York: AIP), in press
\medskip
\noindent
Paczy\'nski, B. 1986, ApJ, 308, L43
\medskip
\noindent
Piran, T. 1992, ApJ, 389, L45
\medskip
\noindent
Press, W., Teukolsky, S., Vetterling, W., \& Flannery, B. 1992,
Numerical Recipes (Cambridge: Cambridge University Press), p. 606
\medskip
\noindent
Rybicki, G., \& Lightman, A. 1979, Radiative Processes in Astrophysics (New
York: Wiley), pp. 140-141
\medskip
\noindent
Shemi, A., \& Piran, T. 1990, ApJ, 365, L55
\medskip
\noindent
Weinberg, S. 1972, Gravitation and Cosmology (New York: Wiley), p. 485
\medskip
\noindent
Wickramasinghe, W. A. D. T., et al. 1993, ApJ, 411, L55
\medskip
\noindent
Usov, V. V. 1992, Nature, 357, 472
\np
\centerline{\bf FIGURE CAPTIONS}

\noindent
{\bf Fig. 1:} Number count $N$ vs.
peak flux $C_{\rm peak}$ for
the 169 bursts in the first BATSE catalog
with $C_{\rm peak}> 1~{\rm cm^{-2}s^{-1}}$ (solid line),
and the best-fit curve for an $\Omega=1$ universe (dotted).

\noindent
{\bf Fig. 2:} Probability distributions $P(x)$ for
the radiation energy $E_{\gamma}$ (Fig. 2a),
the minimum Lorentz factor $\gamma_{\rm min}$ (Fig. 2b),
the maximum baryonic mass $M_{\rm max}$ (Fig. 2c), and
the upper bound on the ambient gas density $n_{\rm max}$ (Fig. 2d).
The histograms show $xP(x)$ for
the variable $x$ whose logarithm appears
on the horizontal axis.
We use
$h=0.75$ and $\epsilon=10(1+z)$ MeV.
Note that only $n_{\rm max}\propto \epsilon^{-7/4}$ is
significantly sensitive to the test photon energy $\epsilon$.

\np

$$ \vbox{\tabskip=0pt \offinterlineskip
\halign{\strut#& \vrule#\tabskip=1em plus2em& \hfil#\hfil&
\vrule#& \hfil#\hfil& \vrule#& \hfil#\hfil& \vrule#& \hfil#\hfil&
\vrule#& \hfil#\hfil& \vrule#\tabskip=0pt\cr
\noalign{\hrule}
\omit&height3pt& & & & & & & &\cr
& & Moment  & & $E_{\gamma}$/erg & & $\gamma_{\rm min}$ & &
$\xi M_{\rm max}/M_{\odot}$ & & $\xi n_{\rm max}/{\rm cm^{-3}}$ & \cr
\noalign{\hrule}
\omit&height3pt& & & & & & & &\cr
& & Mean & & $4.3\times 10^{51}$ & & $4.7\times 10^2$ & & $0.98\times 10^{-5}$
&
& $0.67\times 10^2$ & \cr
\omit&height3pt& & & & & & & &\cr
\noalign{\hrule}
\omit&height3pt& & & & & & & &\cr
& & Std. Dev. & & $7.8\times 10^{51}$ & & $2.2\times 10^2$ & & $2.4\times
10^{-5}$ & & $1.2\times 10^2$ & \cr
\omit&height3pt& & & & & & & &\cr
\noalign{\hrule}
\omit&height3pt& & & & & & & &\cr
& & Skewness & & 3.9 & & 0.70 & & 4.2 & & 3.8 & \cr
\omit&height3pt& & & & & & & &\cr
\noalign{\hrule}  }}$$
\medskip
\noindent
{\bf Table 1:} Mean, standard deviation, and
skewness of the distributions
given in figure 2.  Values of $\xi n_{\rm max}$ are shown for
$\gamma=\gamma_{\rm min}$.
\end